# A cortical sparse distributed coding model linking mini- and macrocolumn-scale functionality

**Gerard J. Rinkus***

*Biology Department, Volen Center for Complex Systems, Brandeis University, Waltham, MA, USA*

****Correspondence:***
*Gerard J. Rinkus, Biology Department, Volen Center for Complex Systems, Brandeis University, 415 South Street, Waltham, MA 02453, USA.*
*e-mail: grinkus@brandeis.edu*

No *generic* function for the minicolumn – i.e., one that would apply equally well to all cortical areas and species – has yet been proposed. I propose that the minicolumn does have a generic functionality, which only becomes clear when seen in the context of the function of the higher-level, subsuming unit, the macrocolumn. I propose that: (a) a macrocolumn's function is to store sparse distributed representations of its inputs and to be a recognizer of those inputs; and (b) the generic function of the minicolumn is to enforce macrocolumnar code sparseness. The minicolumn, defined here as a physically localized pool of ~20 L2/3 pyramidals, does this by acting as a winner-take-all (WTA) *competitive module*, implying that macrocolumnar codes consist of ~70 active L2/3 cells, assuming ~70 minicolumns per macrocolumn. I describe an algorithm for activating these codes during both learning and retrievals, which causes more similar inputs to map to more highly intersecting codes, a property which yields ultra-fast (immediate, first-shot) storage and retrieval. The algorithm achieves this by adding an amount of randomness (noise) into the code selection process, which is inversely proportional to an input's familiarity. I propose a possible mapping of the algorithm onto cortical circuitry, and adduce evidence for a neuromodulatory implementation of this familiarity-contingent noise mechanism. The model is distinguished from other recent columnar cortical circuit models in proposing a generic minicolumnar function in which a group of cells *within* the minicolumn, the L2/3 pyramidals, compete (WTA) to be part of the sparse distributed macrocolumnar code.

**Keywords: sparse distributed representations, minicolumn, macrocolumn, novelty detection, population coding, learning, memory, winner-take-all**

## INTRODUCTION

The columnar organization of neocortex at the minicolumnar (20–50 μm) and macrocolumnar (300–600 μm) scales has long been known (see Mountcastle, 1997; Horton and Adams, 2005 for reviews). Minicolumn-scale organization has been demonstrated on several anatomical bases (Lorente de No, 1938; DeFelipe et al., 1990; Peters and Sethares, 1996). There has been substantial debate as to whether this highly regular minicolumn-scale structure has some accompanying generic dynamics or functionality. See Horton and Adams (2005) for a review of the debate. However, thus far no such *generic* function for the minicolumn – i.e., one that would apply equally well to all cortical areas and species – has been determined.

One basis upon which a functionality for the minicolumn has been suggested is possession of highly similar receptive field characteristics, or tuning, by the cells comprising the minicolumn, e.g., V1 orientation columns (Hubel and Wiesel, 1962, 1968) and minicolumn-sized regions innervating cutaneous zones (Favorov and Diamond, 1990). The reasoning here appears to be that because a group of cells all have very similar tuning to a particular feature, α, e.g., an edge at a particular orientation, they form a unit whose function is to recognize α. However, in searching for a possible generic minicolumn function, we need not limit ourselves to considering only recognition functions. Furthermore, possession of highly similar tuning cannot be a basis for a *generic* minicolumn functionality since in many cortical areas, the cells encountered along vertical penetrations can have quite variable tuning (cf. Sato et al., 2007, 2008). On closer analysis, this is true in orientation column data as well (Hetherington and Swindale, 1999; Ringach et al., 2002). If there is a generic minicolumn functionality, then it must be compatible with varying degrees of tuning correlation amongst the minicolumn's cells.

I propose that the minicolumn does have a generic functionality (given shortly), but one that only becomes clear when seen in the context of the function of the higher-level, subsuming unit, namely, the macrocolumn, which has been demonstrated anatomically (Goldman and Nauta, 1977; Lübke et al., 2003; Egger et al., 2008) and functionally (Mountcastle, 1957; Woolsey and Van der Loos, 1970; Hubel and Wiesel, 1974; Albright et al., 1984). I propose that the function of a macrocolumn (e.g., hypercolumn, segregate, barrel) is to store *sparse distributed representations* of its *overall* input patterns, and to act as a recognizer of those patterns. By "overall input" pattern, I mean the macrocolumn's overall input at a given moment, including not only its bottom-up (BU) inputs from thalamus or lower cortical areas, but also its top-down (TD) inputs from higher cortical areas and its horizontal (H) inputs, which I propose carry temporal (sequential) context information (recurrently) from the representations previously active in the same and nearby macrocolumns. Thus, an "overall input pattern" can equally well be termed, a "context-dependent input". Thus, it is in fact the macrocolumn whose generic functionality is appropriately characterized as recognition; i.e., recognition of a class determined by the history of context-dependent inputs to which it has been exposed.

A distributed representation of an item of information is one in which multiple units collectively represent that item, and crucially, such that each of those units generally participates in the representations of other items as well. Distributed representations are also referred to as cell assemblies, population codes, or ensembles. In this paper, "representation" and "code" will be used interchangeably. A *sparse* distributed code (SDC) is one in which only a small fraction of the pool of available representing units is part of any particular code (Palm, 1982; Lynch et al., 1986; Kanerva, 1988).

If the macrocolumn stores SDCs, then there must be some mechanism that enforces sparseness and this, I propose, is the generic function of the minicolumn. Specifically, I propose that small, physically localized pools of L2/3 pyramidals, e.g., ~20 such cells, act as winner-take-all (WTA) *competitive modules* (CMs). This implies that macrocolumnar codes should consist of about 60–80 active L2/3 cells, one per minicolumn: for simplicity, assume 70 minicolumns per macrocolumn hereafter. Defined in this way, the minicolumn has a more flexible relation to the ontogenetic column, the apical dendrite bundle, the DBC horsetail, etc. For example, a given minicolumn might include L2/3 pyramidals from more than one apical dendrite bundle, consistent with the findings of Yoshimura and Callaway (2005) of fine-scale networks of preferentially interconnected L2/3 pyramidals.

There is increasing evidence for the use of SDC in the cortex and other brain structures; e.g., auditory cortex (Hromdka et al., 2008), visual areas (Young and Yamane, 1992; Vinje and Gallant, 2000; Waydo et al., 2006; Quian Quiroga et al., 2008), zebra finch neopallium (Hahnloser et al., 2002), olfactory structures (Jortner et al., 2007; Linster and Cleland, 2009; Poo and Isaacson, 2009), and hippocampus (Leutgeb et al., 2007). Particularly supportive of the proposed hypothesis is the Reid Lab's calcium imaging data of rat V1 during stimulation by drifting square-wave gratings (Ohki et al., 2005). Their movie (http://reid.med.harvard.edu/movies.html) shows sparse collections of L2/3 cells extending over an approximately macrocolumn-sized region synchronously turning on and off in response to particular grating orientations. **Figures 1A,B** (two frames from the movie) show distinct sets of cells, i.e., codes, responding to bars moving left and right, respectively, and emphasize that individual units may participate in multiple codes (red-circled cells). In terms of the proposed model, the active neurons would be the winners in their respective minicolumns, as suggested in **Figures 1C,D**. **Figure S1** in Supplementary Material provides another view of how the proposed model maps onto cortex.

If the macrocolumn does indeed function as a SDC field in the way suggested here, then we must answer two key questions regarding its dynamics.

1. How is any particular set of winners, one in each of the 70 minicolumns, initially chosen in response to an input pattern and bound into a holistic code? That is, how are macrocolumnar codes *learned*?
2. How is a previously learned code reactivated in response to future presentations of the input pattern that it was initially chosen to represent? That is, how are stored macrocolumnar codes *retrieved* (reactivated)?

In the next section, I describe an algorithm, referred to as the *code selection algorithm* (CSA), which answers both questions. A key property of the CSA is that it causes similar inputs to be assigned to similar, i.e., more highly intersecting, codes. This property, which will be referred to as SISC (similar inputs map to similar codes), is very important in terms of computational efficiency (see Discussion) and is possessed by most distributed coding schemes. However, the CSA achieves it in a novel, probabilistic fashion, which can be summarized as follows:

1. Determine the *familiarity* of a macrocolumn's input. To a first approximation, an input's familiarity is its maximum similarity to any input previously stored in the macrocolumn.
2. Set the amount of randomness (noise) in the process of selecting winners in the WTA modules in inverse proportion to the input's familiarity.
3. Select the winners.

The algorithm's rationale is described in detail in the next section, but broadly, the idea is as follows. When an input, $\alpha_1$, is familiar, we want to reactivate the code, $\beta_1$, to which $\alpha_1$ was previously assigned. The cells comprising $\beta_1$ will have had their synapses (from the cells comprising $\alpha_1$) increased during learning. Thus, if $\alpha_1$ is presented again, the cells of $\beta_1$ will have the highest synaptic input summations in their respective WTA modules. In this case, winners should be chosen on the *deterministic* basis of these summations: no noise should be present in the winner selection process. On the other hand, increasingly novel inputs should be assigned to increasingly distinct codes, i.e., codes having progressively smaller intersection with existing codes. This can be achieved by increasing the noisiness of the winner selection process in each WTA module, which can be achieved by suppressing the influence of the deterministic synaptic inputs (which reflect prior learning) relative to baseline random (spontaneous) activity. By adjusting parameters that control the global (i.e., across the whole macrocolumn) noise level, we can modulate the expected intersection between the set of cells which have the maximal input summations in their respective WTA modules and the set of winners that are actually chosen, thus implementing SISC.

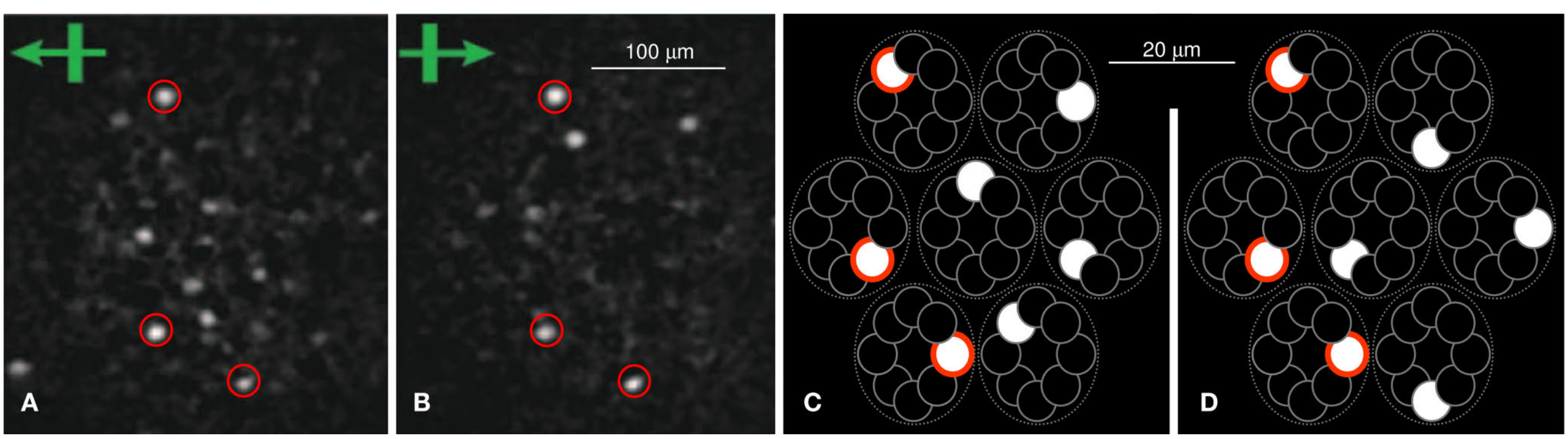

**FIGURE 1 | Calcium (tangential) images of L2/3 of rat visual cortex showing sparse sets of cells activating in unison (see movie link in text) in response to leftward (A) and rightward (B) drifting gratings.** From Ohki et al. (2005). Red circles highlight some cells common to both codes. **(C,D)** Sketch of proposed sparse distributed coding concept, which could plausibly give rise to data like **(A)** and **(B)**. Note different scales. Red borders emphasize intersections between codes. N.b.: To make the sketches look more like the calcium images, black is used for inactive units and white for active: this is reversed in subsequent figures.

Many experimental and theoretical studies implicate neuromodulators, notably norepinephrine (NE) and acetylcholine (ACh), in functionality similar to the above, which can be described generally as modulating signal-to-noise ratio (SNR). Doya (2002) proposed that NE levels control the amount of noise in a process of choosing output actions. However, Doya's model assumes a *localist* representation of the choices, which precludes possession of the SISC property (see Discussion). In addition, increased ACh has been shown to selectively increase the strength of afferent relative to intrinsic inputs in piriform cortex (Hasselmo and Bower, 1992) and other brain structures (see Hasselmo, 2006 for review). These ACh findings have been summarized as showing that increased ACh adjusts network dynamics to favor encoding new memories compared to retrieving old memories, which fits well with the proposed CSA functionality. Following the model description, I offer a speculative mapping of my proposed model onto neural circuitry and discuss evidence for novelty-contingent noise modulation by both NE and ACh. However, the specifics of this mechanism are a subject of ongoing research and likely will involve interactions between neuromodulatory systems (cf. Briand et al., 2007).

Any discussion of columnar function of course centrally concerns cortical circuitry, and more specifically, the putative canonical cortical microcircuit (Rockland and Pandya, 1979; Douglas et al., 1989; Douglas and Martin, 1991). I therefore want to finish the Introduction with the following point. We have made huge progress in understanding many of the components of cortical microcircuitry – a tiny sample of which includes (DeFelipe et al., 1990; Larkum et al., 2001; Beierlein et al., 2003; Schubert et al., 2003; Zhu et al., 2004; Feldmeyer et al., 2006; Fukuda et al., 2006; Krieger et al., 2007; Egger et al., 2008; Hirata and Castro-Alamancos, 2008; Berger et al., 2009; Murayama et al., 2009; Symes and Wennekers, 2009; Briggs, 2010; and see Thomson et al., 2002; Bannister, 2005; Silberberg et al., 2005 for reviews). Nevertheless, we remain far from any sort of comprehensive and consensual understanding of how cortical columnar circuitry manipulates, i.e., stores and retrieves, *specific* items of information. In the main, only very broad conclusions regarding information processing are asserted in the experimental literature, e.g., horizontal connections between fast-spiking L4 interneurons and pyramidals are involved in formation of L4 assemblies and sharpening of tuning (Sun et al., 2006); that both the populations of L4 pyramidals and of L5A pyramidals have transcolumnar connectivity patterns allowing them to act as integrators of information coming in from multiple vibrissae in parallel, or in close sequence (Schubert et al., 2007); that the receptive fields of barrel-related L2/3 pyramids are dynamic and thus may reflect learning to recognize spatiotemporal patterns of vibrissae deflections (Brecht et al., 2003); that WTA competition occurs in the supra- and infragranular layers (Douglas and Martin, 2004); and that local (~100 μm) L2/3-to-L2/3 connections might serve to synchronize firing of L2/3 cell assemblies (Lübke and Feldmeyer, 2007); etc. I believe the hypothetical model described herein to be a significant contribution because it goes beyond broad conclusions and offers a mechanistic explanation of how *specific* informational items are learned and retrieved and in so doing, proposes a generic function for the minicolumn, i.e., that it functions as a WTA module in support of manipulating SDCs at the next higher, i.e., macrocolumnar, scale.

## RESULTS: MODEL DESCRIPTION

**Figure 2** shows the functional architecture of a *simplified* version of the model. In particular, it was stated in the Introduction that sparse macrocolumnar codes are chosen in response to a macrocolumn's *overall* input, which includes its BU, H, and TD inputs. However, illustrations of the model in operation in that general case become rather complex, particularly since the H (and TD) weights carry temporal information, which necessitates showing the model at multiple successive time steps while processing spatiotemporal patterns. More importantly, the core elements of the hypothesis – which are: (a) that the macrocolumn stores SDCs consisting of one winning L2/3 cell per minicolumn; and (b) that the SISC property is achieved by modulating the amount of randomness (noise) present in the winner selection process in inverse proportion to input familiarity – can be clearly and more simply described for the BU-only case (i.e., where inputs are purely spatial patterns). Therefore, the model description in this paper will be limited to the BU-only case. However, the generalized model (with BU, H, and TD inputs) and the accompanying generalized version of the CSA are given in **Figures S2 and Table S1** in Supplementary Material.

In **Figure 2**, the input field, F1, is comprised of 12 binary units and can be considered analogous to a specific thalamic nucleus, though topographical organization is not assumed. The coding field, F2, consists of $Q = 4$ WTA CMs, each containing $K = 3$ binary units. Complete (all-to-all) BU connectivity from F1 to F2 is assumed for simplicity. These BU weights (synapses) are binary, initially 0, and are permanently set to a weight of 1 the first time the pre- and postsynaptic units are co-active (i.e., Hebbian learning).

### MODEL DYNAMICS: THE CODE SELECTION ALGORITHM

The model's core algorithm, the CSA, determines which cells are chosen to represent an input, during both learning and retrieval. A single iteration of the algorithm involves two rounds of competition in the CMs of F2. The first round is a *hard WTA* competition (represented by boxes labeled "Max" in **Figure 2**). The purpose of the first round is to compute a *global familiarity* measure, $G$, of the input pattern. $G$ then drives a global modulation of the F2 unit activation function (**Figure 2**: purple arrows) in preparation for the second competitive round, which is a *soft WTA* competition, the intent of which is that:

1. as $G$ goes to 1 (indicating a completely familiar input), the probability that the unit with the highest input summation in a CM wins approaches 1, and
2. as $G$ goes to 0 (indicating a completely novel input), all units in a CM become equally likely to win (regardless of their input summations).

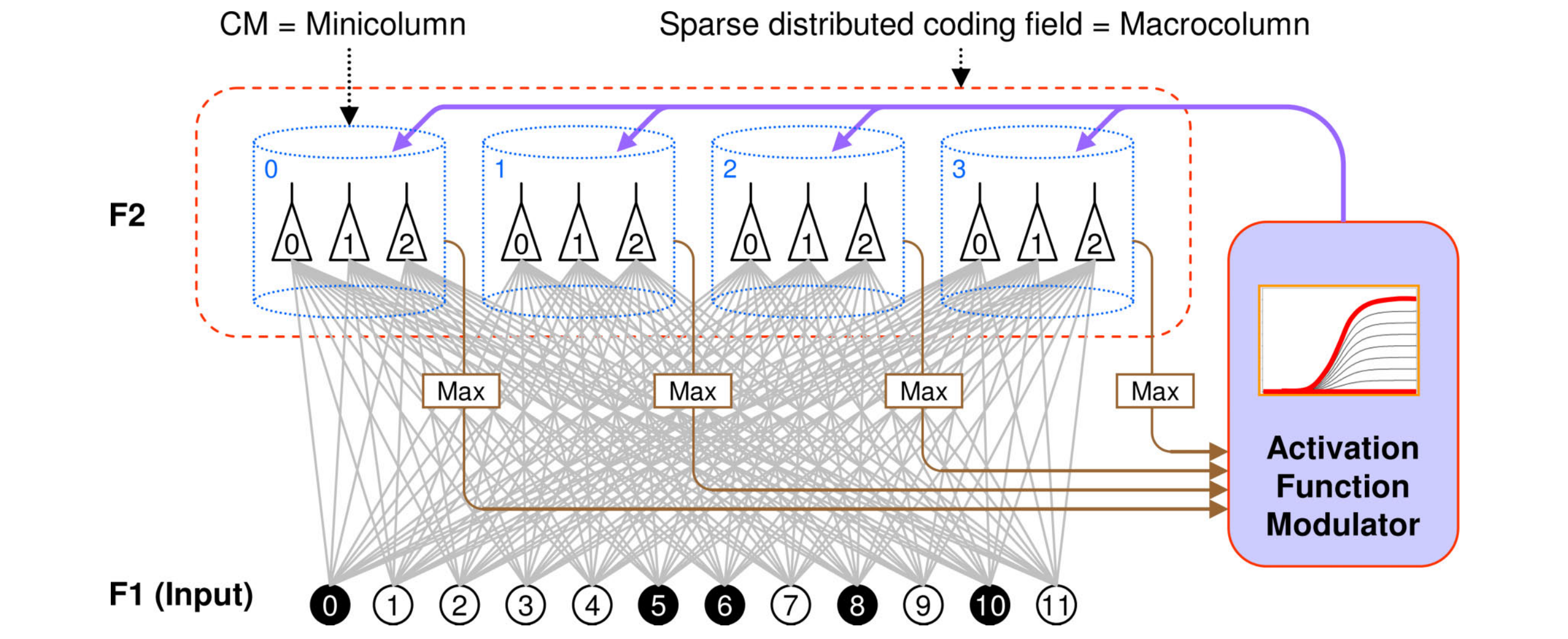

**FIGURE 2 | Functional architecture.** The input field (F1) consists of binary feature detectors: a particular input consisting of five active features is shown. The coding field, F2, is proposed as a macrocolumn analog. It consists of winner-take-all competitive modules (CMs) proposed as analogous to minicolumns. Each CM has $K = 3$ binary units. Bottom-up connectivity is all-to-all: gray lines signify initially 0 weights. The familiarity ($G$) of the input is proposed to be computed via a subcortical, neuromodulator-based mechanism, which then modulates the F2 unit activation function parameters (e.g., sigmoid height) contingent on $G$ (purple arrows). See text.

This policy ensures, statistically, the SISC property. The steps of the CSA are as follows.

1. Each F2 unit $i$ computes its raw input summation, $u(i)$.

$$u(i) = \sum_{j \in \alpha_n} w(j,i) \tag{1}$$

where $\alpha_n$ is the current input (F1) pattern. Because unit activations are binary, we can simply sum the weights, $w(j,i)$, which are also binary.

2. Normalize $u(i)$ to [0..1], yielding $V(i)$.

$$V(i) = u(i)/S \tag{2}$$

$S$ is the number of active units in the input pattern. $V(i)$ is a local measure of support, or likelihood, that F2 unit $i$ should be activated. It reflects how well unit $i$'s receptive field (RF), specified by its afferent weight vector, matches the current input vector.

3. (Round 1 competition) The maximum $V(i)$, $\hat{V}_x$, is found in each of the $Q$ CMs.

$$\hat{V}_x = \max_{i \in C_x} \{V(i)\} \tag{3}$$

where $x$ indexes the CMs and $i$ indexes the units in a CM, $C_x$.

4. Average the $Q$ $\hat{V}_x$ values, yielding $G$, a *global* measure of the *familiarity* of the current input.

$$G \equiv \sum_{x=1}^{Q} \hat{V}_x \Big/ Q \tag{4}$$

5. The *expansivity*, η, of the probabilistic activation function (which is implemented via steps 6–8) is set as an increasing nonlinear function of $G$ (Eq. 5, expressed as a table).

| G | 0.0 | 0.2 | 0.4 | 0.6 | 0.8 | 1.0 |
|---|---|---|---|---|---|---|
| η | 0 | 0 | 0.2 | 5 | 12 | 100 |

(5)

η corresponds to sigmoid height (in Eq. 6). The idea is to increase the range of *relative win likelihoods*, ψ($i$) (defined in step 6) over any given CM's units as $G$ goes to 1. This in turn, serves to nonlinearly exaggerate the difference in the *final win probabilities* (Eq. 7) between F2 units with low and high $V$ values. The specific parameters of any instance of the $G$-to-η mapping will determine the specifics of the relation between input similarity and code similarity, i.e., the expected code intersection as a function of input similarity. The specific η values in Eq. 5 were chosen to yield the ρ-distributions in the examples of **Figures 3 and 4**.

6. The $V$ values of all units in all CMs are then passed through the sigmoidal activation function (Eq. 6) whose shape/scale reflects $G$. Again, particular parameter values affect the relation of input similarity to code similarity (and therefore, storage capacity): values of $\lambda = 28$ and $\phi = -5$ produce the $V$-to-ψ mappings in **Figure 4**. As noted above, within each CM, the output variable, ψ($i$), can be viewed as a relative likelihood that unit $i$ should be chosen winner. The ψ-distributions in each CM are normalized to final probabilities in step 7.

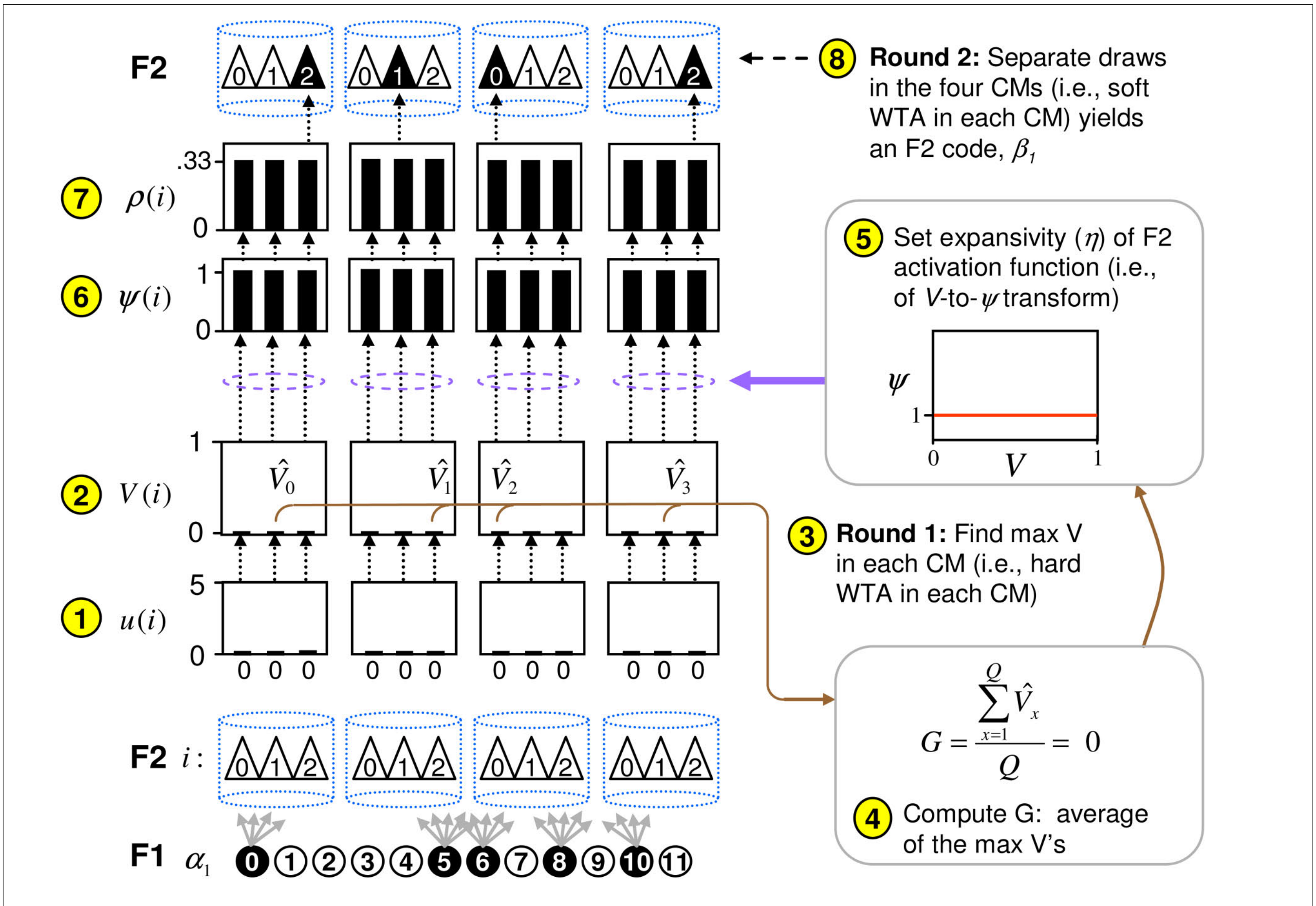

**FIGURE 3 | Graphic illustration of the CSA.** Circled numbers refer to algorithm steps in text. Here, I show the case of the first input, $\alpha_1$, presented to the model. The algorithm detects that $\alpha_1$ is completely unfamiliar, i.e., *G* is computed to be 0, and sets the F2 activation function to a constant function (red line), which makes the choice of winner in each CM completely random, and thus, the overall F2 code, $\beta_1$, also completely random.

$$\psi(i)=\frac{\eta}{1+e^{-(\lambda V(i)+\phi)}}+1 \tag{6}$$

When $G = 1$ (perfectly familiar), η is maximized (in Eq. 5), which maximizes relative and total (once normalized, via Eq. 7) probabilities of winning for units with the maximum *V* value in their respective CMs. In contrast, when $G = 0$ (completely novel), $\eta = 0$, which collapses the sigmoid to the constant function, $\psi = 1$, thus making all units in a CM equally likely to win. This causes the expected intersection of the code being chosen in the current instance with any previously assigned code to be at chance level. In general, this modulation of the sigmoid activation function tends toward *code completion* in proportion to the familiarity of the input and *code separation* in proportion to its novelty.

7. Transform relative likelihood distribution (ψ) in each CM to true probability distribution (ρ).

$$\rho(i)=\frac{\psi(i)}{\sum_{k\in \text{CM}}\psi(k)} \tag{7}$$

8. (Round 2 competition) Choose an F2 code by drawing a winner from the ρ-distribution (soft max) in each CM. Thus, choosing an F2 code is actually performed as *Q* separate draws. When $G = 0$, these draws are statistically independent, as in **Figures 3 and 4D**. As we consider increasingly familiar inputs, i.e., for *G* approaching 1 (and, assuming the model is still operating in a regime where crosstalk is sufficiently low), the draws become increasingly correlated (dependent), as can be seen in going from **Figure 4C** to **4B** to **4A**.

**Figure 3** graphically illustrates the operation of the CSA in the case of the model being presented with its first input, $\alpha_1$. The gray arrows indicate that the BU signals propagating from the active F1 units will be traversing connections with zero synaptic strength. This leads to unnormalized (*u*) and normalized (*V*) input summations of 0 for all 12 F2 units (steps 1,2). In step 3, the max $V, \hat{V}$, in each CM is found (ties broken at random). In step 4, *G* is computed as the average of the $\hat{V}$ values: in this case all the $\hat{V}$ are 0, so $G = 0$. In step 5, the value, $G = 0$, maps to $\eta = 0$, which causes the activation function of the F2 units to collapse to the constant function, $\psi = 1$. In step 6, each F2 unit applies this activation function to its *V* value, yielding the uniform relative likelihood distribution in each CM. In step 7, the relative likelihood function in each CM is normalized to a true probability (ρ) distribution, which in this case, is again uniform. Finally, in step 8, a winner is drawn in each CM, resulting in a random F2 code, e.g., $\beta_1$.

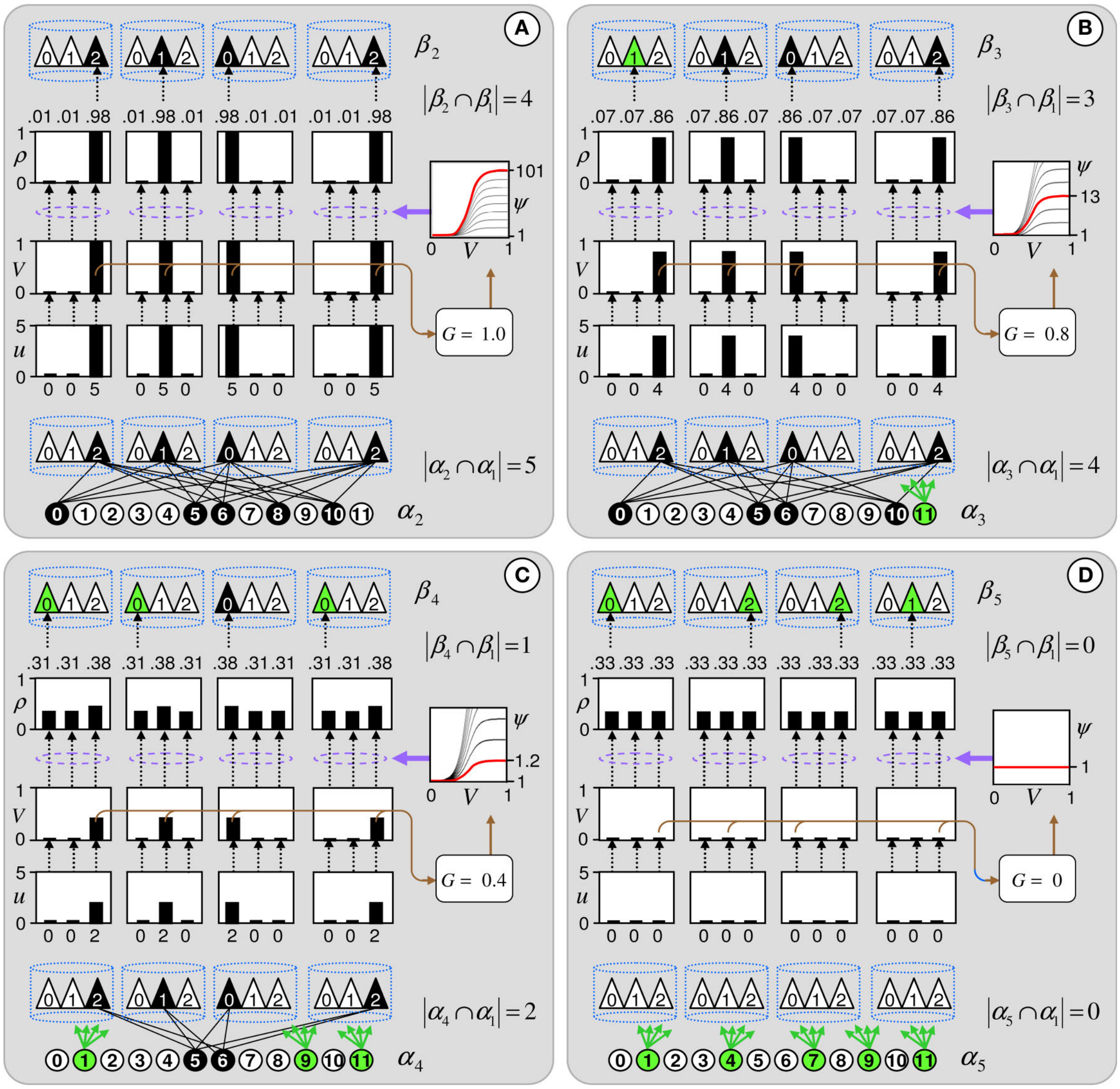

**FIGURE 4 | The CSA and the SISC property.** Green F1 (F2) units denote units *not* in common with $\alpha_1$ ($\beta_1$). Green arrow sprays represent signals propagating via naïve ($w = 0$) weights. See text.

**Figure 4** demonstrates that the CSA realizes the SISC property by considering four possibilities (A–D) for the second input presented to the model of **Figure 3**. These four inputs, $\alpha_2$–$\alpha_5$, range from being identical to $\alpha_1$ (completely familiar) to having zero overlap with $\alpha_1$ (completely unfamiliar). To save space, the panels of **Figure 4** use an abbreviated version of the format of **Figure 3**. Most noticeably, the intermediate variable, $\psi$ (relative likelihood), is not shown. However, the transform from $V$ through to $\rho$ should still be clear. Black BU connections are ones that were increased to one when $\alpha_1$ was learned (**Figure 3**). The overall message of **Figure 4** is as follows. Working from **Figure 4A** to **4D**, the inputs have progressively lower similarity (intersection) with $\alpha_1$: F1 units *not* in common with $\alpha_1$ are shown in green. As $G$ drops, the sigmoid expansivity drops (note the changing $\psi$ scale). Thus, the $\rho$-distributions become progressively flatter, which in turn results in F2 codes, $\beta_2$–$\beta_5$, having progressively smaller intersection with $\beta_1$. F2 units *not* in common with $\beta_1$ also shown in green.

**Figure 4A** shows the case of presenting a completely familiar input again, and is thus a recognition test trial, demonstrating retrieval. This leads, via CSA steps 3 and 4, to $G = 1$, which yields, via steps 5 and 6, the expansive nonlinear $V$-to-$\psi$ mapping shown (red sigmoid). This nonlinearity is applied to every F2 unit, yielding the highly peaked $\rho$-distributions shown. Finally, one unit is drawn in each CM. The probability of drawing the correct unit in any single CM is approximately 98%. Of course, what's crucial in this case, i.e., when the input is completely familiar ($G = 1$), is that the *entire* correct F2 code is reactivated. In this case, that probability is $(0.98)^4 \approx 92\%$. Thus, the familiarity, $G$, which depends on the entire F2 layer and is thus *global* information, influences the *local* activation functions so as to produce the desired overall result, in this case, reactivation of the code (memory trace), $\beta_1$, of the familiar input pattern, $\alpha_1$. The explanations of the remaining panels follow that of **Figures 3 and 4A**. In going from **Figure 4B** to **4D**, one can readily see decreasing intersection with $\alpha_1$, decreasing *u* and *V* values, decreasing *G*, decreasing sigmoid expansivity, progressively flatter ρ-distributions, and ultimately, decreasing intersection with $\beta_1$.

Given a desired probability, *R*, of correctly reactivating an entire code (i.e., of choosing the correct unit in each CM), when *G* = 1, given values for *Q* and *K*, we could compute the needed value of η. However, the macrocolumn model is a content-addressable associative memory and in actual usage, multiple sparse codes will be stored in superposition. Any thorough analysis of the model's expected retrieval error must include the effect of overlap between the stored codes (i.e., cross-talk): this influences the shapes of the ρ-distributions and thus, the expected retrieval accuracy for any given number of stored codes. Such an analysis will be conducted empirically and reported in a separate paper.

Before leaving **Figure 4**, I underscore three important points. First, while the ρ-distributions become flatter as *G* decreases, the units comprising the code of the most similar previously learned input (here, $\alpha_1$) remain most likely to win in their respective CMs. If we simply *deterministically* chose the unit with maximum *V*(*i*) in each CM, we would have chosen the same F2 code, $\beta_1$, in response to all four inputs, $\alpha_2$–$\alpha_5$. Thus, the computation of a quantity, *G*, which depends on *all* the CMs is essential to achieving the SISC property. It constitutes a channel through which information transfers between *all* F2 units throughout the whole macrocolumn. As noted earlier, the full model also assumes direct "H" connections between F2 units, analogous to the horizontal matrix of L2/3 (see **Figure S2** in Supplementary Material). These also mediate communication, but of the prior code active in the macrocolumn, not of the simultaneous state of all F2 units throughout the macrocolumn.

Second, learning is *single-trial* and involves only one iteration of the CSA. This is largely facilitated by the fact that when a given input-code association, $\alpha_j$–$\beta_j$, is learned, each of $\beta_j$'s F2 units simultaneously has its afferent weight from *all* of $\alpha_j$'s F1 units increased. The effect of these simultaneous correlated potentiations allows a rapid, even single-trial, formation of an association, even if the individual synaptic potentiations are small, consistent with the recent characterization of thalamocortical learning described in Bruno and Sakmann (2006).

Third, **Figure 4A** shows that recognizing an exact instance of a previous input also requires only one iteration of the CSA. Although this example does not directly show it, this holds for recognition of non-exact matches as well. Evidence for this will be presented in a separate work. That both learning and recognition require only a single CSA iteration is especially significant since, as can readily be seen, none of the CSA steps involves iterations over stored codes: thus, the time it takes for the CSA to either store a new input or retrieve the closest matching stored input remains constant as the number of stored codes increases. This does not imply that an infinite number of codes can be stored: of course, the model has finite storage capacity. This capacity will be characterized in future research, but should be similar to other sparse associative memories (Willshaw et al., 1969; Palm, 1982; Moll and Miikkulainen, 1995; Knoblauch et al., 2010).

## PROSPECTIVE MAPPING TO CORTICAL CIRCUITRY

There remain huge gaps in our knowledge of the intrinsic physiological, synaptic, morphological, and connectional properties of all classes of cortical cell and of functional relationships between cortical and sub-cortical structures. Nevertheless, **Figure 5** shows a possible neural realization of the model's WTA CM, i.e., minicolumn, and dynamics (CSA). I have attempted to respect known constraints but the realization is admittedly speculative and ongoing modifications will undoubtedly be required. **Figures 5A–E** show the critical events transpiring in a single minicolumn at five points in time during the CSA's computational cycle. Note that due to space limitations **Figure 5** cannot depict the true extents of the various axonal and dendritic systems of the cells involved. **Figure S3** in Supplementary Material provides a more global context showing these extents.

**Figure 5A** shows the initial CSA steps 1 and 2 when the L2/3 pyramidals (here only two cells, but in reality, ~20) integrate their inputs. While the CSA explanation in the prior section included only the BU inputs, all three input classes are included here:

- BU inputs (labeled "2") are mediated via L4 (Rockland and Pandya, 1979)
- TD inputs ("1") are mediated via L2/3 apical tufts (Rockland and Drash, 1996)
- H inputs ("3") are mediated via the horizontal matrix of L2/3 (Gilbert and Wiesel, 1989; Feldmeyer et al., 2006)

All three input vectors arrive in parallel and collectively give rise to postsynaptic potentials (PSPs) in the L2/3 pyramidals. The three (normalized) inputs are combined multiplicatively; see the generalized version of the CSA (**Table S1** in Supplementary Material). The chandelier cells (represented by a single red unit labeled "C") are firing at this time, preventing the L2/3 pyramids from firing.

In **Figure 5B**, the chandeliers shut off (grayed out) and the L2/3 pyramid with the highest PSP (cell 2) is assumed to be the first to spike (CSA step 3). This winning cell, and more specifically, its PSP value (*V* in Eq. 2), represents this minicolumn's *local* judgment of how similar the macrocolumn's closest-matching stored input is to the current overall (i.e., BU, H, and TD) input. The output of cell 2 is then communicated to some locus where it is averaged with the round 1 winners from the other ~70 minicolumns of the macrocolumn, yielding *G* (CSA step 4). As noted in the Introduction, the functionality related to *G* seems most consistent with the phenomenology of neuromodulatory systems, especially ACh and NE. Support for this speculation is given in the following sub-section. Note that the communication of cell 2's PSP value is hypothesized to be mediated via L5, one of whose pyramidal cells is seen integrating here (light green); this is based loosely on evidence that L5 cells, specifically L5B pyramidals, almost exclusively target pontine areas (with collaterals to thalamus) (Deschênes et al., 1994; Krieger et al., 2007).

L2/3 pyramidals are the primary source of BU signals to higher cortical areas (Rockland and Pandya, 1979; and see Thomson et al., 2002; Brecht et al., 2003; Schubert et al., 2003; Bannister, 2005;

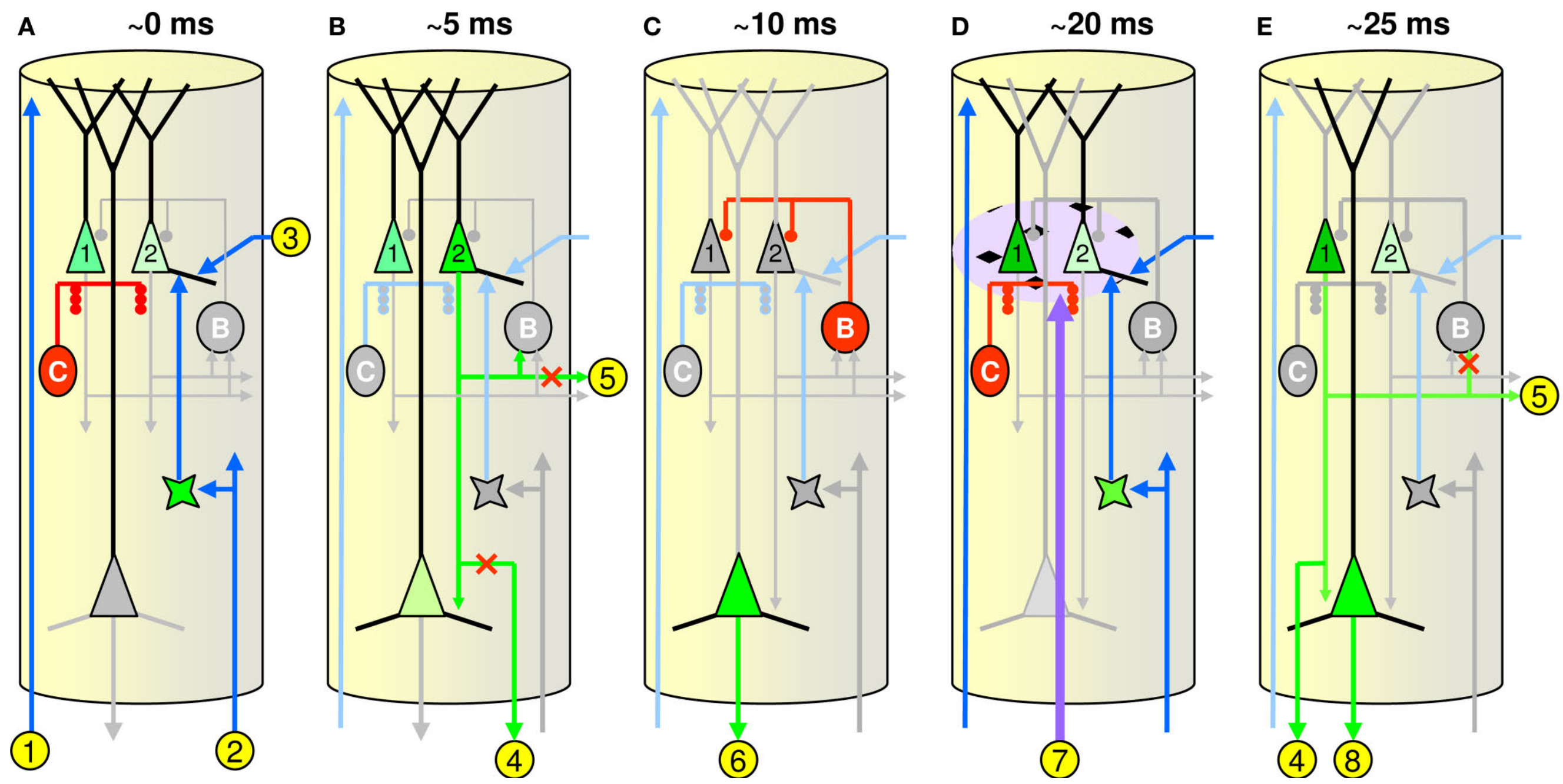

**FIGURE 5 | Hypothetical minicolumnar circuit. (A–E)** Depict critical events transpiring in a single minicolumn at five points in time during the CSA's computational cycle, which I propose corresponds to one gamma cycle: approximate timings relative to start of CSA cycle are shown across top. The units labeled "C" ("B") represent the local chandelier (basket cell) populations, respectively; See text for detailed explanation.

Helmstaedter et al., 2007; Lübke and Feldmeyer, 2007; Petersen, 2007; Egger et al., 2008; Lefort et al., 2009 for wider background on cortical columnar circuitry relevant to the current proposal). In addition, as stated earlier, the horizontal L2/3-to-L2/3 connections are proposed to communicate this macrocolumn's *final* hypothesis regarding its total input pattern in the current CSA cycle recurrently back to the same (and surrounding) macrocolumns on the next CSA cycle. Hence, it is crucial that since that final hypothesis is not present until the second round of competition completes (**Figure 5E**), the output pathways carrying those signals must be closed (red "x"s on paths "4" and "5"). Though not depicted here, one possible mechanism for selectively preventing horizontal signaling in L2/3 is activation of the double bouquet cells (DeFelipe et al., 1990, 2006; Peters and Sethares, 1997). Their "horsetail" axons, being interdigitated, nearly one-to-one with minicolumns would allow them to make contact with a substantial portion of the horizontally (and obliquely) oriented dendritic and axonal processes, in L2/3, and thus prevent passage of horizontal signals.

In **Figure 5C**, the L5 pyramidal mediating the winning L2/3 cell's PSP value has reached threshold and sends that output to the sub-cortical averaging locus (path "6"). In addition, the winning cell itself has activated the local basket cell network (electrically coupled, cf. Brown and Hestrin, 2009), represented by the unit labeled "B", which rapidly deactivates and re-polarizes (resets) the L2/3 pyramidals (grayed out). This reset need not be back to a completely even baseline: some remnant of the relative PSP distribution prior to basket cell activation might remain, biasing the second round of competition.

In **Figure 5D**, the result of the subcortical computation of *G* is returned to the macrocolumn (path "7") in the form of neuromodulator release (purple cloud surrounding the L2/3 pyramidals). This release modifies the activation functions of the L2/3 pyramidals, as described earlier. Note that this neuromodulatory "cloud" actually extends across a broad, i.e., macrocolumnar (or wider), expanse of L2/3, not just within a single minicolumn as this figure may suggest. The chandeliers are again firing to prevent any L2/3 from firing as the second round of integration occurs. The basket cells are inactive, allowing this integration to take place.

In **Figure 5E**, the chandeliers again deactivate. The L2/3 pyramidal with the highest PSP is the first to spike. In general, the pyramidal cell winning on this second round of competition may differ from the first round winner. In particular, the probability that the second round winner is the same as the first round winner increases with *G*. The set of L2/3 winners, one per minicolumn, across the whole macrocolumn represents that macrocolumn's final decision (hypothesis) as to identity of its current overall (TD, H, and BU) input. Thus, the output of the winning L2/3 cell in each minicolumn is now communicated to:

1. Lower cortical regions via L5 and its backprojections to the lower regions' L1 (labeled "8") (Rockland and Drash, 1996).
2. L2/3 pyramids in the same and neighboring macrocolumns via the local horizontal L2/3 matrix ("5") (Gilbert and Wiesel, 1989; Feldmeyer et al., 2006), thus delivering temporal context information (recurrently) to the local region to be integrated on the next CSA cycle.
3. The L4 layer of higher cortical regions ("4") (Rockland and Pandya, 1979).

Note that the output of the winning L2/3 cell should be prevented from recurring to the local basket network at this time so as to allow the integration period to occur at the beginning of the next computational cycle; hence, the red "x" on the link to basket cell.

I reiterate that the above possible cortical realization of the proposed SDC model is highly speculative. It clearly lacks numerous details, especially regarding processing in the intermediate processing stages, e.g., L4, and output processing involving L5 (and L6). Nevertheless, it is a starting point and can be falsified, especially as experimental methods mature. For example, the many timing relationships in the circuit can be tested. We still have decidedly little in the way of hard constraints on the time courses of activation of the many cell types involved in cortical (and hippocampal) circuits, though progress is being made (Klausberger et al., 2003, 2004; Silberberg et al., 2005; Silberberg and Markram, 2007; Klausberger and Somogyi, 2008; Otsuka and Kawaguchi, 2009; Woodruff et al., 2009).

Moreover, the proposed theory's key assumption that the L2/3 pyramidals are the core repository of information in cortex and that the codes laid down in L2/3 are the context-dependent memories of our experiences, is subject to challenge. Specifically, the anatomy of the L5 thick tufted cells suggests that they too have access to BU (via L4), TD (via their apical tufts in L1), and H (via an extensive intra-L5 horizontal network, Schubert et al., 2007) inputs, and therefore that L5 might store the most detailed and context-dependent codes in cortex, a view supported by findings such as (de Kock et al., 2007). In the end, for the purpose of this "hypothesis and theory" paper, I believe the architecture and algorithm (CSA) to be more important than the specifics of any particular neural realization.

### *Support for neuromodulator-based implementation of familiarity-contingent noise*

In this section, I consider evidence relating to six model predictions:

a} *Signals generated in the macrocolumn [i.e., the $\hat{V}_x$ (Eq. 3)] can influence neuromodulatory systems (brown links in* ***Figure 2****).* Strictly interpreted, **Figure 2** suggests that the model can only be true of cortical areas that have direct projections to the activation function modulator (AFM), hypothesized to be instantiated in midbrain neuromodulator source areas, e.g., basal forebrain (BF, source of ACh) and locus coeruleus (LC, source of NE). Relatively few cortical areas project directly to BF or LC. Direct cortical afferents to BF arise mainly from prepyriform, anterior insula, orbitofrontal, entorhinal and medial temporal areas (Mesulam and Mufson, 1984). Direct cortical afferents to LC arise from dorsal prefrontal cortex (Arnsten and Goldman-Rakic, 1984), medial prefrontal cortex (Jodo et al., 1998).

While direct projections are limited, a much larger fraction of cortex may be able to influence the hypothesized AFM via multi-synaptic pathways involving thalamus and other subcortical structures, especially via pathways interconnecting BF, LC, and other neuromodulator source areas. For example, BF cholinergic neurons are excited by LC (Jones et al., 2004), which allows dorsal and medial prefrontal areas indirect influence on BF. Similarly, LC receives input from the Raphe nuclei (reviewed in Samuels and Szabadi, 2008) which would allow further extension of the sphere of cortical influence upon the AFM.

This is a subject of ongoing research. However, it is clearly possible that my macrocolumnar model applies only to the smaller set of cortical areas suggested above. There is some merit to this idea. After all, there would be some advantage to deferring the decision as to the overall familiarity/novelty of the current input (moment) to the very highest cortical levels, which are in position to make the most informed decisions. In this case, once $G$ is computed, it is then broadcast pan-cortically, i.e., to all levels of the hierarchy, allowing the local choice of code to proceed accordingly, i.e., with a level of randomness appropriate to $G$. **Figure S2** in Supplementary Material illustrates this view.

b} *There exists some neuromodulatory signal, $\eta$ (Eq. 5), which correlates with the detection of familiarity, and/or inversely with the detection of novelty.* Such correlations have been shown for both ACh (Miranda et al., 2000, 2003) and NE (Sara et al., 1994; Vankov et al., 1995).

c} *The signal $\eta$ can physically reach cortex (purple arrows in* ***Figure 2****).* LC projects to all of cortex (Foote and Morrison, 1987; Berridge and Waterhouse, 2003; Samuels and Szabadi, 2008). BF projects to almost all cortical areas (reviewed in Briand et al., 2007). The amount of randomness to be added to the winner selection process is a global parameter, which applies *non-specifically* to all the minicolumns. This is consistent with volume transmission believed to be used by neuromodulatory systems (Descarries et al., 1997; see Sarter et al., 2009 for discussion of the complexities of the evidence regarding volume transmission).

d} *The signal $\eta$ determines (Eqs 6–8) the amount of noise (randomness) in the selection (activation) of cortical (i.e., macrocolumnar) codes.* Controlling the noisiness of a process of choosing a winner from a competing group of neurons can be achieved by some combination of two actions: (i) increasing the spike probability of cells with high input summations relative to those with low summations and (ii) lowering the spike probability of low-input cells relative to high-input cells. Both of these actions can be achieved by uniformly (i.e., to all competing cells in a WTA module) modulating intrinsic cell properties such as excitability. Numerous studies have demonstrated both excitatory and suppressive effects on target cell responses (both principal neurons and interneurons) for both ACh (Kawasaki and Avoli, 1996; Shalinsky et al., 2002; Cobb and Davies, 2005; Tateno et al., 2005; Isakova and Mednikova, 2007; Lawrence, 2008; Eggermann and Feldmeyer, 2009) and NE (Foote et al., 1975; Kawaguchi and Shindou, 1998; Harley and Helen, 2007; Moxon et al., 2007). It is not my intention here to argue for a precise realization of this mechanism. As suggested in many reviews (Berridge and Waterhouse, 2003; Lucas-Meunier et al., 2003; Sara, 2009), the landscape of this research is very complex and we are far from a comprehensive understanding of the how the various neuromodulatory systems affect high-level cognitive processing either alone or in concert (Briand et al., 2007). Nevertheless, the large and varied body of evidence at least admits the possibility that one or more of these neuromodulators could implement the familiarity-contingent AFM mechanism (CSA steps 5–8; see Doya, 2002, p. 502, for similar speculation).

e} *Decreased* $\eta$, *i.e., increased noise, leads to greater code separation (decreased intersection).* Recently, Goard and Dan (2009) showed that increased BF stimulation decreased the correlation amongst a population of rat V1 cells. This decreased correlation essentially shows increased separation between population codes, which in the model proposed here, would manifest as decreased intersection between sparse codes.

f} *Disabling of the brain's ability to produce high noise, i.e., causing* $\eta$ *to be permanently high, should reduce the ability to learn new inputs, while sparing or having much less effect on recognition of known items.* Looking at **Figure 4**, if the AFM was "stuck" in the highly expansive sigmoid condition (low noise), all four inputs, $\alpha_2$–$\alpha_5$, would have high probability of mapping to the same code, $\beta_1$. This would prevent the model from being able to distinguish them in future presentations. However, in general, inputs that were mapped to unique codes prior to such a disabling event will reliably activate those codes on future presentations. In accord with this, Browning et al. (2010) found that severely diminishing cholinergic inputs to inferotemporal cortex severely reduced macaques' performance on a visual episodic memory task, while having little effect on a DNMS task. McGaughy et al. (2005) found a similar effect: cholinergic deafferentation of entorhinal cortex reduced performance on DNMS tasks involving novel odors but not familiar odors.

## DISCUSSION

I have described a theoretical model of cortical function that explains the functional relationship between the minicolumn and macrocolumn. Specifically:

a} The macrocolumn (in any of its forms) is proposed to store information in the form of SDCs, and

b} The minicolumn (specifically, its L2/3 pool of pyramidals) is proposed to operate as a WTA CM, the purpose of which is to enforce the sparseness of the macrocolumnar code.

Two key motivations for this model are the computational advantages of SDC and the increasingly strong evidence for SDC in the brain, cited in the Section "Introduction". One important advantage of SDC over a localist code is that the number of unique items that can be stored can be far larger than the number of representing units. For example, the 12 F2 units of the model in **Figure 2** allow $3^4 = 81$ unique codes, though in realistic systems, e.g., with less than complete connectivity leading to and from a coding field like F2, the number of those codes that can safely (i.e., while maintaining retrieval error rates below some acceptable criterion) be assigned will be substantially lower than 81. Nevertheless, if the number of input items that will need to be distinguished is not known *a priori*, SDC is more flexible.

A second computational advantage of SDC is that, if used in conjunction with an appropriate storage/retrieval algorithm it possesses the SISC property. I demonstrated, with the small but statistically reasonable example of **Figures 3 and 4**, that the CSA yields the SISC property. The SISC property strongly differentiates SDC from localist models: it is not even defined for a localist model since every code is formally disjoint with every other code. Hence, there is no structural way to represent degrees of similarity in a localist code. If there is no way to represent a measure, e.g., similarity, structurally, then whenever that measure is required – e.g., when the closest matching stored item in a database (i.e., macrocolumn) to an input must be returned – it must be computed, which takes time and energy. In contrast, when items' codes are stored in physically overlapped fashion such that the degree of code overlap represents item similarity, as is the case for the proposed model, the most closely matching stored item will be returned *immediately*, i.e., without requiring any serial search through the stored items. **Figure S4** in Supplementary Material shows test retrievals of the four unique codes stored in the model of **Figures 3 and 4**, demonstrating possession of this immediate access property for this small example. Empirical proof of this property based on larger scale simulations is currently being developed.

I consider the representation and the CSA to be the most important contributions of this paper because of the computational advantages just described. However, I believe the suggestion that the minicolumn's *generic* function is to act as a WTA CM is also important. Saying only that a group of L2/3 units forms a WTA CM places no *a priori* constraints on what their tuning functions or receptive fields should look like. This is what gives that functionality a chance of being truly generic, i.e., of applying across all areas and species, regardless of the observed tuning profiles of closely neighboring units. Indeed, a recent calcium imaging study of mouse auditory cortex by Rothschild et al. (2010) shows highly heterogeneous small-scale (even immediately adjacent cells) tuning even though the large-scale tuning is tonotopic. Experimental methods are only now just reaching the point where this hypothesis might be directly testable, e.g., modifying calcium imaging methods to have millisecond temporal granularity; see Ohki and Reid (2007).

In a sense, the main point of this paper is that a *generic* minicolumnar function becomes apparent as soon as we postulate that what the cortex, i.e., a macrocolumn, generally does is store and retrieve (access) SDCs of *specific* context-dependent inputs. As noted in the Section "Introduction", the experimental literature contains little in the way of proposals linking the formation and retrieval of *specific* SDCs (i.e., of specific input items, especially of temporal context-dependent items) to the cortical microcircuitry. My proposed model goes beyond broad conclusions and offers a mechanistic explanation of how *specific* informational items are learned and retrieved and in so doing, proposes a generic function for the minicolumn, i.e., that it functions as a WTA module in support of manipulating SDCs at the next higher, i.e., macrocolumnar, scale.

There have been several recent models linking formation/retrieval of specific items to cortical circuitry and which describe specific roles for the minicolumn (Kupper et al., 2007; George and Hawkins, 2009; Litvak and Ullman, 2009; Schrader et al., 2009). However, all of these models use localist representations and therefore would not possess the advantages of SDC described above. The Cortext model (Kupper et al., 2007; Schrader et al., 2009) assumes that each minicolumn in a hypercolumn represents one particular input feature. On each computational cycle, a WTA process runs within each *hypercolumn*, such that exactly one minicolumn wins, which would be strongly at odds with the calcium image data (Ohki et al., 2005). A second problem is that the assumption that whole minicolumns compete with each other implies that any given hypercolumn (at any level of the cortical hierarchy) can represent only ~70 unique features (equivalence classes), which seems severely restrictive, especially for hypercolumns at higher cortical levels, e.g., IT. The Litvak and Ullman (2009) model also postulates that the L2/3 pool of neurons in a minicolumn implements a max function. However, their model proposes that each single minicolumn (specifically, its L2/3 pool) is partitioned into several *disjoint* groups ("cliques") of cells, each representing a different item. Since any particular cell can participate in only one clique, this constitutes a localist code. George and Hawkins (2009) also assume that minicolumns represent informational items in a localist fashion. Note however that both George and Hawkins (2009) and Litvak and Ullman (2009) explicitly mention moving to a more general combinatorial code, a.k.a. SDC, as a future research direction.

A core principle of the proposed model is this notion of controlling the amount of noise in the process of choosing (activating) a macrocolumnar code as an inverse function of input similarity. Doya (2002) uses the same principle, referred to as "Boltzmann selection", to modulate the amount of noise in the process of choosing amongst output action actions. Doya specifically hypothesizes that NE controls the noise whereas I can assert only that it is some neuromodulator-based mechanism. In Doya's model, as NE levels increase, the action with the greatest expected reward is chosen with probability approaching 1. This is suggested as corresponding to the "exploitation" end of the exploitation–exploration continuum. As NE levels drop to 0, all actions become equally probable, i.e., "exploration", which is appropriate if no single action has a particular high expected reward, which generally correlates with the condition of novelty, i.e., of being in a novel environment wherein it is harder to anticipate the outcome of known actions. The analogy to high expected reward in my model is a highly familiar input ($G \approx 1$) in which case we want the stored code for that familiar input to be reactivated with probability approaching 1; the condition where no action has a high expected reward is analogous to low familiarity, i.e., where no stored input is very similar to the current input, in which case we want to lay down a new memory trace for the novel input. Despite the similarities, Doya's model also assumes a *localist* representation of the choices and, like the other models just mentioned, cannot realize the advantages of SDC.

I have identified several avenues of active and future research at various points in the text and as noted in the previous section, the prospective neural realization is highly speculative and very incomplete. Several additional questions/issues for future research are:

1. Is the current proposal that the L2/3 cells engage in two rounds of competition in each computational (putatively, gamma) cycle plausible?
2. For simplicity, I have described the model in the simplest case of having only one internal coding field (F2) and processing only purely spatial input patterns. However the core model was originally developed for the spatiotemporal pattern (sequence) case (Rinkus, 1996) and was generalized some time ago to have an arbitrarily deep hierarchy of coding fields (Rinkus and Lisman, 2005). See **Figure S2** in Supplementary Material. How do these generalized versions of the model map to neural structures?
3. Is there evidence that chandeliers become active twice as frequently as baskets, as the proposed realization predicts? Is there evidence for the converse?
4. Although not elaborated herein, the proposed mini-/macrocolumn model is easily generalized to allow substantial overlap between minicolumns (see **Figure S5** in Supplementary Material) and multiple winners in a minicolumn on each computational cycle. These degrees of freedom need to be explored.
5. We know of the fast, phasic, time scales of operation for NE (Rajkowski et al., 2004) and DA (Schultz, 1998) and of slightly slower but still phasic mode for ACh (Gulledge and Kawaguchi, 2007), but these have been proposed as signaling other measures besides pure novelty (Redgrave et al., 1999; Bouret and Sara, 2005; Dayan and Yu, 2006). How might all these signals conspire to implement a pure novelty signal?

## ACKNOWLEDGMENTS

This work was partially supported by NIH Grant, 5 T32 NS07292. I thank John Lisman as well the reviewers for many valuable suggestions and insights that have greatly abetted the development of these ideas.

## SUPPLEMENTARY MATERIAL

The Supplementary Material for this article can be found online at http://www.frontiersin.org/neuroscience/neuroanatomy/paper/10.3389/fnana.2010.00017/

## REFERENCES

Albright, T. D., Desimone, R., and Gross, C. G. (1984). Columnar organization of directionally selective cells in visual area MT of the macaque. *J. Neurophysiol.* 51, 16–31.

Arnsten, A. F. T., and Goldman-Rakic, P. S. (1984). Selective prefrontal cortical projections to the region of the locus coeruleus and raphe nuclei in the rhesus monkey. *Brain Res.* 306, 9–18.

Bannister, A. P. (2005). Inter- and intralaminar connections of pyramidal cells in the neocortex. *Neurosci. Res.* 53, 95–103.

Beierlein, M., Gibson, J. R., and Connors, B. W. (2003). Two dynamically distinct inhibitory networks in layer 4 of the neocortex. *J. Neurophysiol.* 90, 2987–3000.

Berger, T. K., Perin, R., Silberberg, G., and Markram, H. (2009). Frequency-dependent disynaptic inhibition in the pyramidal network: a ubiquitous pathway in the developing rat neocortex. *J. Physiol.* 587, 5411–5425.

Berridge, C. W., and Waterhouse, B. D. (2003). The locus coeruleus-noradrenergic system: modulation of behavioral state and state-dependent cognitive processes. *Brain Res. Rev.* 42, 33.

Bouret, S., and Sara, S. J. (2005). Network reset: a simplified overarching theory of locus coeruleus noradrenaline function. *Trends Neurosci.* 28, 574–582.

Brecht, M., Roth, A., and Sakmann, B. (2003). Dynamic receptive fields of reconstructed pyramidal cells in layers 3 and 2 of rat somatosensory barrel cortex. *J. Physiol. (Lond.)* 553, 243–265.

Briand, L. A., Gritton, H., Howe, W. M., Young, D. A., and Sarter, M. (2007). Modulators in concert for cognition: modulator interactions in the prefrontal cortex. *Prog. Neurobiol.* 83, 69–91.

Briggs, F. (2010). Organizing principles of cortical layer 6. *Front. Neural Circuits.* 4:3. doi: 10.3389/neuro.04.003.2010.

Brown, S. P., and Hestrin, S. (2009). Cell-type identity: a key to unlocking the function of neocortical circuits. *Curr. Opin. Neurobiol.* 19, 415–421.

Browning, P. G. F., Gaffan, D., Croxson, P. L., and Baxter, M. G. (2010). Severe scene learning impairment, but intact recognition memory, after cholinergic depletion of inferotemporal cortex followed by fornix transection. *Cereb. Cortex* 20, 282–293.

Bruno, R. M., and Sakmann, B. (2006). Cortex is driven by weak but synchronously active thalamocortical synapses. *Science* 312, 1622–1627.

Cobb, S. R., and Davies, C. H. (2005). Cholinergic modulation of hippocampal cells and circuits. *J. Physiol.* 562, 81–88.

Dayan, P., and Yu, A. J. (2006). Phasic norepinephrine: a neural interrupt signal for unexpected events. *Netw. Comput. Neural Syst.* 17, 335–350.

DeFelipe, J., Ballesteros-Yáñez, I., Inda, M. C., and Muñoz, A. (2006). Double-bouquet cells in the monkey and human cerebral cortex with special reference to areas 17 and 18. *Prog. Brain Res.* 154 (Pt. 1), 15–32.

DeFelipe, J., Hendry, S. H. C., Hashikawa, T., Molinari, M., and Jones, E. G. (1990). A microcolumnar structure of monkey cerebral cortex revealed by immunocytochemical studies of double bouquet cell axons. *Neuroscience* 37, 655–673.

de Kock, C. P. J., Bruno, R. M., Spors, H., and Sakmann, B. (2007). Layer- and cell-type-specific suprathreshold stimulus representation in rat primary somatosensory cortex. *J. Physiol.* 581, 139–154.

Descarries, L., Gisiger, V., and Steriade, M. (1997). Diffuse transmission by acetylcholine in the CNS. *Prog. Neurobiol.* 53, 603–625.

Deschênes, M., Bourassa, J., and Pinault, D. (1994). Corticothalamic projections from layer V cells in rat are collaterals of long-range corticofugal axons. *Brain Res.* 664, 215–219.

Douglas, R. J., and Martin, K. A. (1991). A functional microcircuit for cat visual cortex. *J. Physiol.* 440, 735–769.

Douglas, R. J., and Martin, K. A. C. (2004). Neuronal circuits of the neocortex. *Annu. Rev. Neurosci.* 27, 419–451.

Douglas, R. J., Martin, K. A., and Witteridge, D. (1989). A canonical microcircuit for neocortex. *Neural. Comput.* 1, 480–488.

Doya, K. (2002). Metalearning and neuromodulation. *Neural Netw.* 15, 495–506.

Egger, V., Nevian, T., and Bruno, R. M. (2008). Subcolumnar dendritic and axonal organization of spiny stellate and star pyramid neurons within a barrel in rat somatosensory cortex. *Cereb. Cortex* 18, 876–889.

Eggermann, E., and Feldmeyer, D. (2009). Cholinergic filtering in the recurrent excitatory microcircuit of cortical layer 4. *Proc. Natl. Acad. Sci. U.S.A.* 106, 11753–11758.

Favorov, O., and Diamond, M. (1990). Demonstration of discrete place-defined columns – segregates – in the cat SI. *J. Comp. Neurol.* 298, 97–112.

Feldmeyer, D., Lübke, J., and Sakmann, B. (2006). Efficacy and connectivity of intracolumnar pairs of layer 2/3 pyramidal cells in the barrel cortex of juvenile rats. *J. Physiol.* (*Lond.*) 575, 583–602.

Foote, S. L., Freedman, R., and Oliver, A. P. (1975). Effects of putative neurotransmitters on neuronal activity in monkey auditory cortex. *Brain Res.* 86, 229–242.

Foote, S. L., and Morrison, J. H. (1987). Extrathalamic modulation of cortical function. *Annu. Rev. Neurosci.* 10, 67–95.

Fukuda, T., Kosaka, T., Singer, W., and Galuske, R. A. W. (2006). Gap junctions among dendrites of cortical GABAergic neurons establish a dense and widespread intercolumnar network. *J. Neurosci.* 26, 3434–3443.

George, D., and Hawkins, J. (2009). Towards a mathematical theory of cortical micro-circuits. *PLoS Comput. Biol.* 5, e1000532. doi: 10.1371/journal.pcbi.1000532.

Gilbert, C., and Wiesel, T. (1989). Columnar specificity of intrinsic horizontal and corticocortical connections in cat visual cortex. *J. Neurosci.* 9, 2432–2442.

Goard, M., and Dan, Y. (2009). Basal forebrain activation enhances cortical coding of natural scenes. *Nat. Neurosci.* 12, 1444–1449.

Goldman, P. S., and Nauta, W. J. H. (1977). Columnar distribution of cortico-cortical fibers in the frontal association, limbic, and motor cortex of the developing rhesus monkey. *Brain Res.* 122, 393–413.

Gulledge, A. T., and Kawaguchi, Y. (2007). Phasic cholinergic signaling in the hippocampus: functional homology with the neocortex? *Hippocampus* 17, 327–332.

Hahnloser, R. H. R., Kozhevnikov, A. A., and Fee, M. S. (2002). An ultra-sparse code underlies the generation of neural sequences in a songbird. *Nature* 419, 65.

Harley, C. W., and Helen, E. S. (2007). Norepinephrine and the dentate gyrus. *Prog. Brain Res.* 163, 299–318.

Hasselmo, M. E. (2006). The role of acetylcholine in learning and memory. *Curr. Opin. Neurobiol.* 16, 710–715.

Hasselmo, M. E., and Bower, J. M. (1992). Cholinergic suppression specific to intrinsic not afferent fiber synapses in rat piriform (olfactory) cortex. *J. Neurophysiol.* 67, 1222–1229.

Helmstaedter, M., de Kock, C. P. J., Feldmeyer, D., Bruno, R. M., and Sakmann, B. (2007). Reconstruction of an average cortical column in silico. *Brain Res. Rev.* 55, 193.

Hetherington, P. A., and Swindale, N. V. (1999). Receptive field and orientation scatter studied by tetrode recordings in cat area 17. *Vis. Neurosci.* 16, 637–652.

Hirata, A., and Castro-Alamancos, M. A. (2008). Cortical transformation of wide-field (multiwhisker) sensory Responses. *J. Neurophysiol.* 100, 358–370.

Horton, J. C., and Adams, D. L. (2005). The cortical column: a structure without a function. *Philos. Trans. R. Soc. Lond., B, Biol. Sci.* 360, 837–862.

Hromdka, T., DeWeese, M. R., and Zador, A. M. (2008). Sparse representation of sounds in the unanesthetized auditory cortex. *PLoS Biol.* 6, e16. doi: 10.1371/journal.pbio.0060016.

Hubel, D. H., and Wiesel, T. N. (1962). Receptive fields, binocular interaction and functional architecture in the cat's visual cortex. *J. Physiol.* 160, 106–154.

Hubel, D. H., and Wiesel, T. N. (1968). Receptive fields and functional architecture of monkey striate cortex. *J. Physiol.* 195, 215–243.

Hubel, D. H., and Wiesel, T. N. (1974). Uniformity of monkey striate cortex: a parallel relationship between field size, scatter, and magnification factor. *J. Comp. Neurol.* 158, 295–305.

Isakova, A., and Mednikova, Y. (2007). Comparative roles of acetylcholine and noradrenaline in controlling the spontaneous activity of cortical neurons. *Neurosci. Behav. Physiol.* 37, 689.

Jodo, E., Chiang, C., and Aston-Jones, G. (1998). Potent excitatory influence of prefrontal cortex activity on noradrenergic locus coeruleus neurons. *Neuroscience* 83, 63–79.

Jones, B. E., Descarries, L., Krnjevic, K., and Steriade, M. (2004). Activity, modulation and role of basal forebrain cholinergic neurons innervating the cerebral cortex. *Prog. Brain Res.* 145, 157–169.

Jortner, R. A., Farivar, S. S., and Laurent, G. (2007). A simple connectivity scheme for sparse coding in an olfactory system. *J. Neurosci.* 27, 1659–1669.

Kanerva, P. (1988). *Sparse Distributed Memory*. Cambridge, MA: MIT Press.

Kawaguchi, Y., and Shindou, T. (1998). Noradrenergic excitation and inhibition of GABAergic cell types in rat frontal cortex. *J. Neurosci.* 18, 6963–6976.

Kawasaki, H., and Avoli, M. (1996). Excitatory effects induced by carbachol on bursting neurons of the rat subiculum. *Neurosci. Lett.* 219, 1–4.

Klausberger, T., Magill, P. J., Marton, L. F., Roberts, J. D. B., Cobden, P. M., Buzsaki, G., and Somogyi, P. (2003). Brain-state- and cell-type-specific firing of hippocampal interneurons in vivo. *Nature* 421, 844–848.

Klausberger, T., Marton, L. F., Baude, A., Roberts, J. D. B., Magill, P. J., and Somogyi, P. (2004). Spike timing of dendrite-targeting bistratified cells during hippocampal network oscillations in vivo. *Nat. Neurosci.* 7, 41–47.

Klausberger, T., and Somogyi, P. (2008). Neuronal diversity and temporal dynamics: the unity of hippocampal circuit operations. *Science* 321, 53–57.

Knoblauch, A., Palm, G., and Sommer, F. T. (2010). Memory capacities for synaptic and structural plasticity. *Neural. Comput.* 22, 289–341.

Krieger, P., Kuner, T., and Sakmann, B. (2007). Synaptic connections between layer 5B pyramidal neurons in mouse somatosensory cortex are independent of apical dendrite bundling. *J. Neurosci.* 27, 11473–11482.

Kupper, R., Knoblauch, A., Gewaltig, M.-O., Körner, U., and Körner, E. (2007). Simulations of signal flow in a functional model of the cortical column. *Neurocomputing* 70, 1711–1716.

Larkum, M. E., Zhu, J. J., and Sakmann, B. (2001). Dendritic mechanisms underlying the coupling of the dendritic with the axonal action potential initiation zone of adult rat layer 5 pyramidal neurons. *J. Physiol.* 533, 447–466.

Lawrence, J. J. (2008). Cholinergic control of GABA release: emerging parallels between neocortex and hippocampus. *Trends Neurosci.* 31, 317–327.

Lefort, S., Tomm, C., Floyd Sarria, J. C., and Petersen, C. C. H. (2009). The excitatory neuronal network of the C2 barrel column in mouse primary somatosensory cortex. *Neuron* 61, 301–316.

Leutgeb, J. K., Leutgeb, S., Moser, M.-B., and Moser, E. I. (2007). Pattern separation in the dentate gyrus and CA3 of the hippocampus. *Science* 315, 961–966.

Linster, C., and Cleland, T. A. (2009). Glomerular microcircuits in the olfactory bulb. *Neural Netw.* 22, 1169–1173.

Litvak, S., and Ullman, S. (2009). Cortical circuitry implementing graphical models. *Neural Comput.* 21, 3010–3056.

Lorente de No, R. (1938). "The cerebral cortex: architecture, intracortical connections, motor projections," in *Physiology of the Nervous System*, ed. J. F. Fulton (London: Oxford University Press), 274–301.

Lübke, J., and Feldmeyer, D. (2007). Excitatory signal flow and connectivity in a cortical column: focus on barrel cortex. *Brain Struct. Funct.* 212, 3–17.

Lübke, J., Roth, A., Feldmeyer, D., and Sakmann, B. (2003). Morphometric analysis of the columnar innervation domain of neurons connecting layer 4 and layer 2/3 of juvenile rat barrel cortex. *Cereb. Cortex* 13, 1051–1063.

Lucas-Meunier, E., Fossier, P., Baux, G., and Amar, M. (2003). Cholinergic modulation of the cortical neuronal network. *Pflugers Arch.* 446, 17–29.

Lynch, G., Shepherd, G. M., Black, I. B., and Killackey, H. P. (1986). *Synapses, Circuits, and the Beginnings of Memory*. Cambridge, MA: MIT Press.

McGaughy, J., Koene, R. A., Eichenbaum, H., and Hasselmo, M. E. (2005). Cholinergic deafferentation of the entorhinal cortex in rats impairs encoding of novel but not familiar stimuli in a delayed nonmatch-to-sample task. *J. Neurosci.* 25, 10273–10281.

Mesulam, M.-M., and Mufson, E. J. (1984). Neural inputs into the nucleus basalis of the substantia innominata (Ch4) in the rhesus monkey. *Brain* 107, 253–274.

Miranda, M. I., Ferreira, G., Ramirez-Lugo, L., and Bermudez-Rattoni, F. (2003). Role of cholinergic system on the construction of memories: taste memory encoding. *Neurobiol. Learn. Mem.* 80, 211–222.

Miranda, M. I., Ramírez-Lugo, L., and Bermúdez-Rattoni, F. (2000). Cortical cholinergic activity is related to the novelty of the stimulus. *Brain Res.* 882, 230–235.

Moll, M., and Miikkulainen, R. (1995). *Convergence-Zone Episodic Memory: Analysis and Simulations*. Technical Report AI95-227. Austin: Department of Computer Science, University of Texas at Austin.

Mountcastle, V. B. (1957). Modality and topographic properties of single neurons of cat's somatic sensory cortex. *J. Neurophysiology.* 20, 408–434.

Mountcastle, V. B. (1997). The columnar organization of the neocortex. *Brain* 120, 701–722.

Moxon, K. A., Devilbiss, D. M., Chapin, J. K., and Waterhouse, B. D. (2007). Influence of norepinephrine on somatosensory neuronal responses in the rat thalamus: a combined modeling and in vivo multi-channel, multi-neuron recording study. *Brain Res.* 1147, 105–123.

Murayama, M., Perez-Garci, E., Nevian, T., Bock, T., Senn, W., and Larkum, M. E. (2009). Dendritic encoding of sensory stimuli controlled by deep cortical interneurons. *Nature* 457, 1137–1141.

Ohki, K., Chung, S., Ch'ng, Y. H., Kara, P., and Reid, R. C. (2005). Functional imaging with cellular resolution reveals precise micro-architecture in visual cortex. *Nature* 433, 597–603.

Ohki, K., and Reid, R. C. (2007). Specificity and randomness in the visual cortex. *Curr. Opin. Neurobiol.* 17, 401–407.

Otsuka, T., and Kawaguchi, Y. (2009). Cortical inhibitory cell types differentially form intralaminar and interlaminar subnetworks with excitatory neurons. *J. Neurosci.* 29, 10533–10540.

Palm, G. (1982). *Neural Assemblies: An Alternative Approach to Artificial Intelligence*. Berlin: Springer.

Peters, A., and Sethares, C. (1996). Myelinated axons and the pyramidal cell modules in monkey primary visual cortex. *J. Comp. Neurol.* 365, 232–255.

Peters, A., and Sethares, C. (1997). The organization of double bouquet cells in monkey striate cortex. *J. Neurocytol.* 26, 779.

Petersen, C. C. H. (2007). The functional organization of the barrel cortex. *Neuron* 56, 339–355.

Poo, C., and Isaacson, J. S. (2009). Odor representations in olfactory cortex: "sparse" coding, global inhibition, and oscillations. *Neuron* 62, 850–861.

Quian Quiroga, R., Kreiman, G., Koch, C., and Fried, I. (2008). Sparse but not "grandmother-cell" coding in the medial temporal lobe. *Trends Cogn. Sci.* 12, 87–91.

Rajkowski, J., Majczynski, H., Clayton, E., and Aston-Jones, G. (2004). Activation of monkey locus coeruleus neurons varies with difficulty and performance in a target detection task. *J. Neurophysiol.* 92, 361–371.

Redgrave, P., Prescott, T. J., and Gurney, K. (1999). Is the short-latency dopamine response too short to signal reward error? *Trends Neurosci.* 22, 146.

Ringach, D. L., Hawken, M. J., and Shapley, R. (2002). Receptive field structure of neurons in monkey primary visual cortex revealed by stimulation with natural image sequences. *J. Vis.* 2, 12–24.

Rinkus, G. (1996). *A Combinatorial Neural Network Exhibiting Episodic and Semantic Memory Properties for Spatio-Temporal Patterns*. Boston, MA: Boston University.

Rinkus, G., and Lisman, J. (2005). Time-invariant recognition of spatiotemporal patterns in a hierarchical cortical model with a caudal-rostral persistence gradient. In *Society for Neuroscience Annual Meeting*, Washington, DC.

Rockland, K. S., and Drash, G. W. (1996). Collateralized divergent feedback connections that target multiple cortical areas. *J. Comp. Neurol.* 373, 529–548.

Rockland, K. S., and Pandya, D. N. (1979). Laminar origins and terminations of cortical connections of the occipital lobe in the rhesus monkey. *Brain Res.* 179, 3–20.

Rothschild, G., Nelken, I., and Mizrahi, A. (2010). Functional organization and population dynamics in the mouse primary auditory cortex. *Nat. Neurosci.* 13, 353–360.

Samuels, E. R., and Szabadi, E. (2008). Functional neuroanatomy of the noradrenergic locus coeruleus: its roles in the regulation of arousal and autonomic function part I: principles of functional organisation. *Curr. Neuropharmacol.* 6, 235–253.

Sara, S. J. (2009). The locus coeruleus and noradrenergic modulation of cognition. *Nat. Rev. Neurosci.* 10, 211–223.

Sara, S. J., Vankov, A., and Hervé, A. (1994). Locus coeruleus-evoked responses in behaving rats: a clue to the role of noradrenaline in memory. *Brain Res. Bull.* 35, 457–465.

Sarter, M., Parikh, V., and Howe, W. M. (2009). Phasic acetylcholine release and the volume transmission hypothesis: time to move on. *Nat. Rev. Neurosci.* 10, 383–390.

Sato, T., Uchida, G., and Tanifuji, M. (2008). Cortical columnar organization is reconsidered in inferior temporal cortex. *Cereb. Cortex* 19, 1870–1888.

Sato, T. R., Gray, N. W., Mainen, Z. F., and Svoboda, K. (2007). The functional microarchitecture of the mouse barrel cortex. *PLoS Biol.* 5, e189. doi: 10.1371/journal.pbio.0050189.

Schrader, S., Gewaltig, M.-O., Körner, U., and Körner, E. (2009). Cortext: a columnar model of bottom-up and top-down processing in the neocortex. *Neural Netw.* 22, 1055–1070.

Schubert, D., Kötter, R., and Staiger, J. (2007). Mapping functional connectivity in barrel-related columns reveals layer- and cell type-specific microcircuits. *Brain Struct. Funct.* 212, 107–119.

Schubert, D., Kötter, R., Zilles, K., Luhmann, H. J., and Staiger, J. F. (2003). Cell type-specific circuits of cortical layer IV spiny neurons. *J. Neurosci.* 23, 2961–2970.

Schultz, W. (1998). Predictive reward signal of dopamine neurons. *J. Neurophysiol.* 80, 1–27.

Shalinsky, M. H., Magistretti, J., Ma, L., and Alonso, A. A. (2002). Muscarinic activation of a cation current and associated current noise in entorhinal-cortex layer-II neurons. *J. Neurophysiol.* 88, 1197–1211.

Silberberg, G., Grillner, S., LeBeau, F. E. N., Maex, R., and Markram, H. (2005). Synaptic pathways in neural microcircuits. *Trends Neurosci.* 28, 541–551.

Silberberg, G., and Markram, H. (2007). Disynaptic inhibition between neocortical pyramidal cells mediated by Martinotti cells. *Neuron* 53, 735–746.

Sun, Q.-Q., Huguenard, J. R., and Prince, D. A. (2006). Barrel cortex microcircuits: thalamocortical feedforward inhibition in spiny stellate cells is mediated by a small number of fast-spiking interneurons. *J. Neurosci.* 26, 1219–1230.

Symes, A., and Wennekers, T. (2009). Spatiotemporal dynamics in the cortical microcircuit: a modelling study of primary visual cortex layer 2/3. *Neural Netw.* 22, 1079–1092.

Tateno, T., Jimbo, Y., and Robinson, H. P. C. (2005). Spatio-temporal cholinergic modulation in cultured networks of rat cortical neurons: spontaneous activity. *Neuroscience* 134, 425–437.

Thomson, A. M., West, D. C., Wang, Y., and Bannister, A. P. (2002). Synaptic connections and small circuits involving excitatory and inhibitory neurons in layers 2–5 of adult rat and cat neocortex: triple intracellular recordings and biocytin labelling in vitro. *Cereb. Cortex* 12, 936–53.

Vankov, A., Herve-Minvielle, A., and Sara, S. J. (1995). Response to novelty and its rapid habituation in locus coeruleus neurons of the freely exploring rat. *Eur. J. Neurosci.* 7, 1180–1187.

Vinje, W. E., and Gallant, J. L. (2000). Sparse coding and decorrelation in primary visual cortex during natural vision. *Science* 287, 1273–1276.

Waydo, S., Kraskov, A., Quian Quiroga, R., Fried, I., and Koch, C. (2006). Sparse representation in the human medial temporal lobe. *J. Neurosci.* 26, 10232–10234.

Willshaw, D. J., Buneman, O. P., and Longuet-Higgins, H. C. (1969). Non holographic associative memory. *Nature* 222, 960–962.

Woodruff, A., Xu, Q., Anderson, S. A., and Yuste, R. (2009). Depolarizing effect of neocortical chandelier neurons. *Front Neural Circuits* 3:15. doi: 10.3389/neuro.04.015.2009.

Woolsey, T. A., and Van der Loos, H. (1970). The structural organization of layer IV in the somatosensory region (S I) of mouse cerebral cortex: the description of a cortical field composed of discrete cytoarchitectonic units. *Brain Res.* 17, 205–242.

Yoshimura, Y., and Callaway, E. M. (2005). Fine-scale specificity of cortical networks depends on inhibitory cell type and connectivity. *Nat. Neurosci.* 8, 1552–1559.

Young, M. P., and Yamane, S. (1992). Sparse population coding of faces in the inferotemporal cortex. *Science* 256, 1327–1331.

Zhu, Y., Stornetta, R. L., and Zhu, J. J. (2004). Chandelier cells control excessive cortical excitation: characteristics of whisker-evoked synaptic responses of layer 2/3 nonpyramidal and pyramidal neurons. *J. Neurosci.* 24, 5101–5108.

**Conflict of Interest Statement:** The author declares that the research was conducted in the absence of any commercial or financial relationships that could be construed as a potential conflict of interest.